\documentclass[english,10pt,final,twocolumn]{elsarticle}
\usepackage[T1]{fontenc}
\usepackage[latin9]{inputenc}
\usepackage{pdfpages}
\usepackage{amstext}
\usepackage{graphicx}
\usepackage{gensymb}
\usepackage{amssymb}
\usepackage{placeins}
\usepackage{subfigure}
\usepackage[title]{appendix}
\makeatletter
\usepackage{float}
\usepackage{multirow}

\usepackage{blindtext}
\usepackage{titlesec}
\usepackage{setspace}

\titleformat{\section}
  {\bfseries\large}{\thesection}{1em}{}
\titleformat{\subsection}
  {\bfseries}{\thesubsection}{1em}{}
  \titleformat{\appendix}
  {\bfseries\large}{\thesection}{1em}{}

\newcommand{\lyxmathsym}[1]{\ifmmode\begingroup\def\b@ld{bold}
  \text{\ifx\math@version\b@ld\bfseries\fi#1}\endgroup\else#1\fi}
\usepackage{braket}
\usepackage{mathrsfs}
\usepackage{slashed}
\usepackage{bm}
\usepackage{datetime}
\usepackage{siunitx}

\usepackage{amsmath}

\DeclareSIUnit[number-unit-product = {}]\clight{c}
\DeclareSIUnit\eVperc{\eV\per\clight}
\DeclareSIUnit\GeVpercs{\giga\eV\squared\per\clight\squared}
\DeclareSIUnit\MeVpercs{\mega\eV\per\clight\squared}
\journal{Physics Letters B}
\biboptions{numbers,sort&compress}
\usepackage[left=1.6cm,right=1.04cm,top=1.65cm,bottom=1.65cm,columnsep=25pt]{geometry}
\usepackage{lineno}
\usepackage[normalem]{ulem}

\usepackage{hyperref}  
\hypersetup{breaklinks = true, colorlinks = true, allcolors = magenta}
\usepackage{setspace}
\usepackage{graphicx}
\usepackage{wrapfig}

\usepackage{xcolor, soul}
\sethlcolor{yellow}

\makeatother

\usepackage{babel}
\begin{document}

\begin{frontmatter}{}

\title{
Measurement of polarization transfer \\in the quasi-elastic $^{40}{\rm Ca}(\vec{e},e' \vec{p})$ process 
}

\author[TAU]{T.~Kolar\corref{cor2}}
\ead{tkolar@mail.tau.ac.il}
\author[Mainz,JLab]{P.~Achenbach}
\author[Mainz]{M.~Christmann}
\author[Mainz]{M.O.~Distler}
\author[Mainz]{L.~Doria}
\author[Mainz]{P.~Eckert}
\author[Mainz]{A.~Esser}
\author[Pavia]{C.~Giusti}
\author[Mainz]{J.~Geimer}
\author[Mainz]{P.~G\"ulker}
\author[Mainz]{M.~Hoek}
\author[Mainz]{P.~Klag}
\author[TAU]{J.~Lichtenstadt}
\author[Mainz]{M.~Littich}
\author[Mainz]{T.~Manoussos}
\author[Mainz]{D.~Markus}
\author[Mainz]{H.~Merkel}
\author[UL,JSI]{M.~Mihovilovi\v{c} }
\author[Mainz]{J.~M\"uller}
\author[Mainz]{U.~M\"uller}
\author[Mainz]{J.~P\"{a}tschke}
\author[UCR]{S.J.~Paul}
\author[TAU]{E.~Piasetzky}
\author[Mainz]{S.~Plura}
\author[Mainz]{J.~Pochodzalla}
\author[JSI]{M.~Po\v{z}un}
\author[huji]{G.~Ron}
\author[Mainz]{B.S.~Schlimme}
\author[Mainz]{C.~Sfienti}
\author[Mainz]{S.~Stengel}
\author[USK]{E.~Stephan}
\author[USC]{S.~Strauch}
\author[Mainz]{C.~Szyszka}
\author[UL,JSI]{S.~\v{S}irca}
\author[Mainz]{M.~Thiel}
\author[USK]{A.~Wilczek}
\author{\\\textbf{(A1 Collaboration)}}

\cortext[cor2]{Corresponding author}
\fntext[JLab]{Present address: Thomas Jefferson National Accelerator Facility, Newport News, VA 23606, USA.}

\address[TAU]{School of Physics and Astronomy, Tel Aviv University, Tel Aviv 69978,
Israel.}
\address[Mainz]{Institut f\"ur Kernphysik, Johannes Gutenberg-Universit\"at, 55099
Mainz, Germany.}
\address[Pavia]{INFN, Sezione di Pavia, via A.~Bassi 6, I-27100 Pavia, Italy.}
\address[UL]{Faculty of Mathematics and Physics, University of Ljubljana, 1000
Ljubljana, Slovenia.}
\address[JSI]{Jo\v{z}ef Stefan Institute, 1000 Ljubljana, Slovenia.}
\address[UCR]{Department of Physics and Astronomy, University of California, Riverside, CA 92521, USA.}
\address[huji]{Racah Institute of Physics, Hebrew University of Jerusalem, Jerusalem
91904, Israel.}
\address[USK]{Institute of Physics, University of Silesia in Katowice, 41-500 Chorz\'ow, Poland.}
\address[USC]{University of South Carolina, Columbia, South Carolina 29208, USA.}

\begin{abstract}
Polarization transfer to a bound proton in polarized electron knock-out reactions, $\mathrm{A}(\vec{e},e^{\prime}\vec{p})$, is a powerful tool to look for an in-medium modification of the bound proton. It requires comparison to calculations that consider the many-body effects accompanying the quasi-free process. We report here measured components $P_x^{\prime}$, $P_z^{\prime}$, and their ratio $P_x^{\prime}/P_z^{\prime}$, of polarization transfer to protons bound in $^{40}\mathrm{Ca}$, which is described well by the shell model and for which reliable calculations are available. While the calculations capture the essence of the data, our statistical precision allows us to observe deviations that cannot be explained by simple scaling, including by varying the proton electromagnetic form factor ratio $G_E/G_M$. We further explore the deviations of the ratio of the polarization transfer components from that of a free proton, $(P_x^{\prime}/P_z^{\prime})_{\rm A}/(P_x^{\prime}/P_z^{\prime})_{\rm H}$, and its dependence on the bound-proton virtuality.
\end{abstract}
\date{\today}

\end{frontmatter}{}

\section{Introduction}
The effect of the nuclear medium on its constituent nucleons is a long-standing question which triggered many hypotheses~\cite{PhysRevC.31.232,BERGMANN1990185,LU1998217,PhysRevC.60.068201,PhysRevC.76.055206,PhysRevC.87.028202} and was followed by experiments which tried to observe differences between free and bound nucleons, mainly protons. Obviously, a straightforward way is to compare a reaction on a free nucleon to that of quasi-elastic scattering on a single bound nucleon. Quasi-elastic scattering is an invaluable tool that gives us insight into the properties of the single bound nucleon, proton wave functions, spectroscopic factors, and more. 

In quasi-elastic kinematics, we are also sensitive to other many body effects, such as final state interactions, meson-exchange currents (MEC), and isobar configurations. The data obtained in many experiments over the years were confronted with calculations which resulted in a better understanding of these effects. However, the question of medium modifications of the nucleon structure when embedded in the nucleus was not resolved. The electromagnetic properties of the nucleon, its charge and magnetization densities, are derived directly from its electromagnetic form factors (EM FFs). A measurement of \textit{effective} EM FFs of a bound nucleon, and comparing it to that of a free one, may point to changes in the EM distributions in the bound nucleons.

Polarized electron beams opened up new possibilities to measure nucleon electromagnetic properties. The ratio of the transverse to longitudinal polarization transfer components in elastic ${}^{1}{\rm H}(\vec{e},e'\vec{p})$ are directly related (in the one photon exchange approximation) to the ratio of the electric and magnetic form factors of the proton $G_E/G_M$~\cite{Akh74}. Despite richer nuclear response, measurements of polarization transfer to a bound proton are also sensitive to the proton EM FFs, and thus, can reveal nuclear medium modifications in the bound nucleon~\cite{PhysRevC.40.290, Kelly:1996hd}. However, distinguishing between the aforementioned many-body effects and in-medium modifications of the proton is not trivial and requires a comparison of the measurements with theoretical calculations that consider the contribution of such many body effects to the quasi-elastic process. Deviations from such calculations (whose input are the free-proton FFs) may be then attributed to changes in the proton electromagnetic structure.

Polarization-transfer in quasi-free proton knock-out has been measured on light nuclei (${}^{2}{\rm H}$~\cite{Milbrath:1997de, PhysRevC.73.064004, deep2012PLB,deepCompPLB,deepPaul}, ${}^{4}{\rm He}$~\cite{Dieterich:2000mu, Strauch, Paolone}, ${}^{12}{\rm C}$~\cite{ceepLet, ceepComp, ceepTim}, and ${}^{16}{\rm O}$~\cite{Malov_O16}), and compared to that on a free proton. The deviations from the free proton ratio for these nuclei, evaluated through the \textit{double ratio} of the polarization transfer components, $(P^{\prime}_x/P^{\prime}_z)_{\rm A}/(P^{\prime}_x/P^{\prime}_z)_{^{1}\mathrm{H}}$, obtained at different kinematics, was shown to be in good agreement when compared at the same virtuality of the struck proton. The analysis comparing the data on protons from the s- and p-shells in ${}^{12}{\rm C}$ to the relativistic distorted-wave impulse approximation (RDWIA) calculation did not show a consistent deviation to suggest in-medium effects~\cite{ceepTim}.

We present here first measurements of the polarization transfer to a bound proton in ${}^{40}{\rm Ca}$, extending previous measurements on light nuclei to a medium size nucleus, which has been studied extensively both experimentally and theoretically~\cite{LAPIKAS1993297,KRAMER1989199,KramerPhD}. The nuclear structure of ${}^{40}{\rm Ca}$ allows us to cleanly separate protons knocked out of $1\mathrm{d}_{3/2}$ and $2\mathrm{s}_{1/2}$ shells and evaluate corresponding polarization transfer components and their ratio. We also study the behavior of the polarization \textit{double ratio} which, in light nuclei, was shown to have a similar dependence on the struck nucleon virtuality independent of the target atomic number. The relatively large range of the proton initial momentum probed in this measurement, and data obtained for protons of different shells in ${}^{40}{\rm Ca}$, are compared to RDWIA calculations to search for possible medium modifications of the FFs.

\section{Experimental setup and kinematics}
The data were acquired at the Mainz Microtron (MAMI) by using two spectrometers of the A1 experimental setup~\cite{a1aparatus}. Spectrometer C was used for the detection of the scattered electron, while spectrometer A was used for protons. The standard detector package of both spectrometers consists of vertical drift chambers (VDCs), followed by two layers of scintillation counters and the \v{C}erenkov radiation detector. In spectrometer A a \v{C}erenkov radiation detector was replaced by a $7\,\mathrm{cm}$ thick carbon analyzer and horizontal drift chambers (HDCs) which, together with VDCs, serve as a focal plane polarimeter (FPP)~\cite{Pospischil:2000pu} used to measure the polarization of knocked-out protons.

\begin{table}[h!]
\caption{
Central kinematic settings of the $^{40}{\rm Ca}(\vec{e},e^{\prime}\vec{p}\,)$ reaction measurement of this work. The number of events passing the event pre-selection for particular $E_{\rm miss}$ range are also given.  
}
\begin{center}
\begin{tabular}{lll}
\hline\hline
\multicolumn{3}{c}{Kinematic setting}  \\
\hline
$E_{\rm beam}$ & [MeV]                   &  600 \\
$Q^2$          & [$({\rm GeV}\!/\!c)^2$] & 0.25\\
$p_{\rm miss}$ & [MeV$\!/\!c$]           & $-200$ to $17$ \\
$p_e$          & [MeV$\!/\!c$]           & 396 \\
$\theta_e$     & [deg]                   & -61.8\\
$p_p$          & [MeV$\!/\!c$]           & 630\\
$\theta_{p}$   & [deg]                   & 40.2\\
\hline\hline
\\
\hline\hline
\multicolumn{3}{c}{\# of events passing cuts ($\times 10^6$)}\\
\hline
All              && 2.43 \\ 
$1{\rm d}_{3/2}$ && 0.18 \\ 
$2{\rm s}_{1/2}$ && 0.23 \\ 
Cont.            && 1.10 \\ 
\hline\hline
\end{tabular}

\end{center}
\label{tab:kinematics}
\end{table}

The $600\,{\rm MeV}$ longitudinally polarized beam was guided on to the $^{40}{\rm Ca}$ target consisting of three $0.41\,{\rm mm}$ thick foils positioned at an angle of $45^{\circ}$ (facing spectrometer A) and $15.0\,{\rm mm}$ from each other (along the beamline). This gives us the maximum luminosity while minimizing the in-target path of the outgoing protons and hence reducing energy losses as well as possible depolarization effects on the knock-out protons. To determine the incident beam polarization and ensure its longitudinal orientation, a M{\o}ller polarimeter was used. As a cross-check Mott polarimetry was performed at the beginning of the experiment and later when the source cathode was replaced. It is a known effect to see polarization increase while the quantum efficiency of the cathode drops, hence we have seen polarization ranging from $79.6\%$ to $88.0\%$. We accounted for having two distinct periods and the variation within each with two independent linear fits of all relevant measurements. The relevant reaction kinematics is shown in Fig.~\ref{fig:kinematics_diagram}. We obtained data in parallel kinematics ($\vec{p}_i\parallel\vec{q}$) summarized in Table \ref{tab:kinematics}.

\section{Analysis}
\begin{figure}[hb!]
\includegraphics[width=\columnwidth]{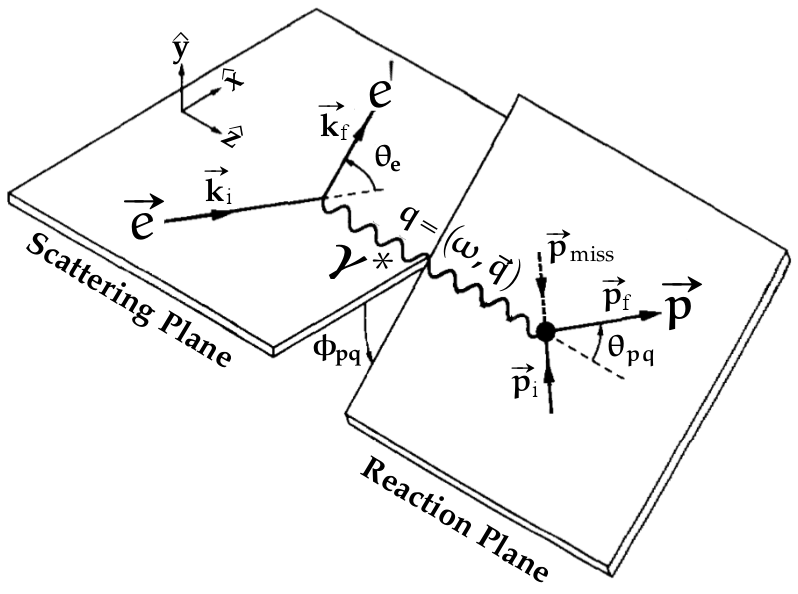}
\caption{
Kinematics of the $(\vec{e},e^{\prime}\vec{p}\,)$ reaction with the definitions of the kinematic variables.
}
\label{fig:kinematics_diagram}
\end{figure}

The polarization-transfer at the reaction point is presented in the $\hat{x},\hat{y},\hat{z}$ coordinate system shown in Fig. \ref{fig:kinematics_diagram} following the convention of~\cite{Strauch}. Both $P^{\prime}_z$ and $P^{\prime}_x$ were determined in the scattering plane, defined by the incident and scattered electron momenta, where $P_z$ is along and $P_x$ is perpendicular to the momentum transfer vector.

We used the FPP to measure the angular asymmetry in secondary scattering of polarized protons due to the spin-orbit part of the NN interaction. Angles are obtained by combining the tracking information from VDCs and HDCs since those correspond to proton tracks before and after the scattering in the analyzer, respectively. Once divided by the polarization independent part, $\sigma_0(\vartheta)$, the angular distribution is given by
\begin{equation}\label{eq:FPPDistro}
\frac{\sigma(\vartheta,\varphi)}{\sigma_0(\vartheta)}=1+A_C(\vartheta,E_{p'})(P_y^{FPP}\cos\varphi-P_x^{FPP}\sin\varphi)\,,
\end{equation}
where $A_C$ is the analyzing power of the carbon scatterer, $\vartheta$ is the polar angle, $\varphi$ is the azimuthal angle. The two transverse polarization components of the proton at the focal plane, $P_x^{\rm FPP}$ and $P_y^{\rm FPP}$, are, depending on the precession of the proton spin in the magnetic field of the spectrometer, function of all three polarization components at target. The relation between two sets of polarization components is described with spin-transfer matrix, $\mathbf{S}$, as
\begin{equation}
\label{eq:reducedSTM}
 \begin{pmatrix}P_x \\ P_y \end{pmatrix}^{\!\mathrm{FPP}} =
    \begin{pmatrix}
        S_{xx} & S_{xy} & S_{xz} \\
        S_{yx} & S_{yy} & S_{yz}  
    \end{pmatrix}\begin{pmatrix}P_x \\ P_y \\ P_z \end{pmatrix}^{\!\mathrm{tg}}\,.
\end{equation} 
Individual components in the reaction plane were obtained by using a well established maximum-likelihood procedure. The statistical uncertainties associated with the extracted components and their ratios have been estimated by using the numerical second-order partial derivative of the log-likelihood function. These uncertainties also include a portion of the systematic spin-transfer error. More details on the procedure can be found in~\cite{ceepTim}.

The polarization components and their ratios have been determined as a function of the proton missing momentum defined as $\vec{p}_{\rm miss}=\vec{q}-\vec{p}_f$. We define the scalar missing momentum, $p_{\rm miss}\equiv\pm |\vec{p}_{\rm miss}|$, where the sign is taken to be positive (negative) if the longitudinal component of $\vec{p}_{\rm miss}$ is parallel (anti-parallel) to $\vec{q}$. 

After careful calibration, certain data cuts were implemented to ensure the quality of the data used in the polarization transfer analysis. First, to identify coincident electrons and protons, a coincidence time cut, $|t_{\rm AC}|<1.5\,\mathrm{ns}$ was made. Second, position resolution at target of $3\to5\,\mathrm{mm}$ allowed us to do a vertex position cut and ensure that the two particles originate from the same target foil. This cut reduced random coincidence background by an order of magnitude to around $10^{-3}$ relative to the coincidence peak. Third, a set of cuts was applied to ensure a good track reconstruction with the drift chambers (e.g. bad wire exclusion, number of wire hits and their distribution between and within layers). Fourth, to separate electrons from other minimum-ionizing particles, we required sufficient energy deposit in threshold \v{C}erenkov detector in spectrometer C. Because of the specific geometry of the detector, this cut also suppressed cosmics background. Fifth, additional cuts were necessary to reliably measure polarization. To ensure secondary scattering occurred in the analyzer, we required the intersect of the two tracks to lie within $\pm 7\,\mathrm{cm}$ from the center of the carbon plate. We also restricted spectrometer A momentum, angular, and position acceptance to the region where magnetic field and hence spin precession is best understood. In effort to eliminate false asymmetries due to the final acceptance and bad wires in HDCs, we performed the ``cone test'', where we only accept events for which their imaginary counterparts with $\varphi^{\rm Test} = \varphi^{\rm FPP} \pm 180^{\circ}$ would be accepted too. Finally, to enhance nuclear over Coulomb scattering contribution, we selected only protons that scattered by more than $10^{\circ}$ in the carbon analyzer. Due to limited knowledge of the carbon analyzing power at much larger angles, we required the maximum scattering angle to be less than $45^{\circ}$. 

\begin{figure}[h]
\includegraphics[width=\columnwidth]{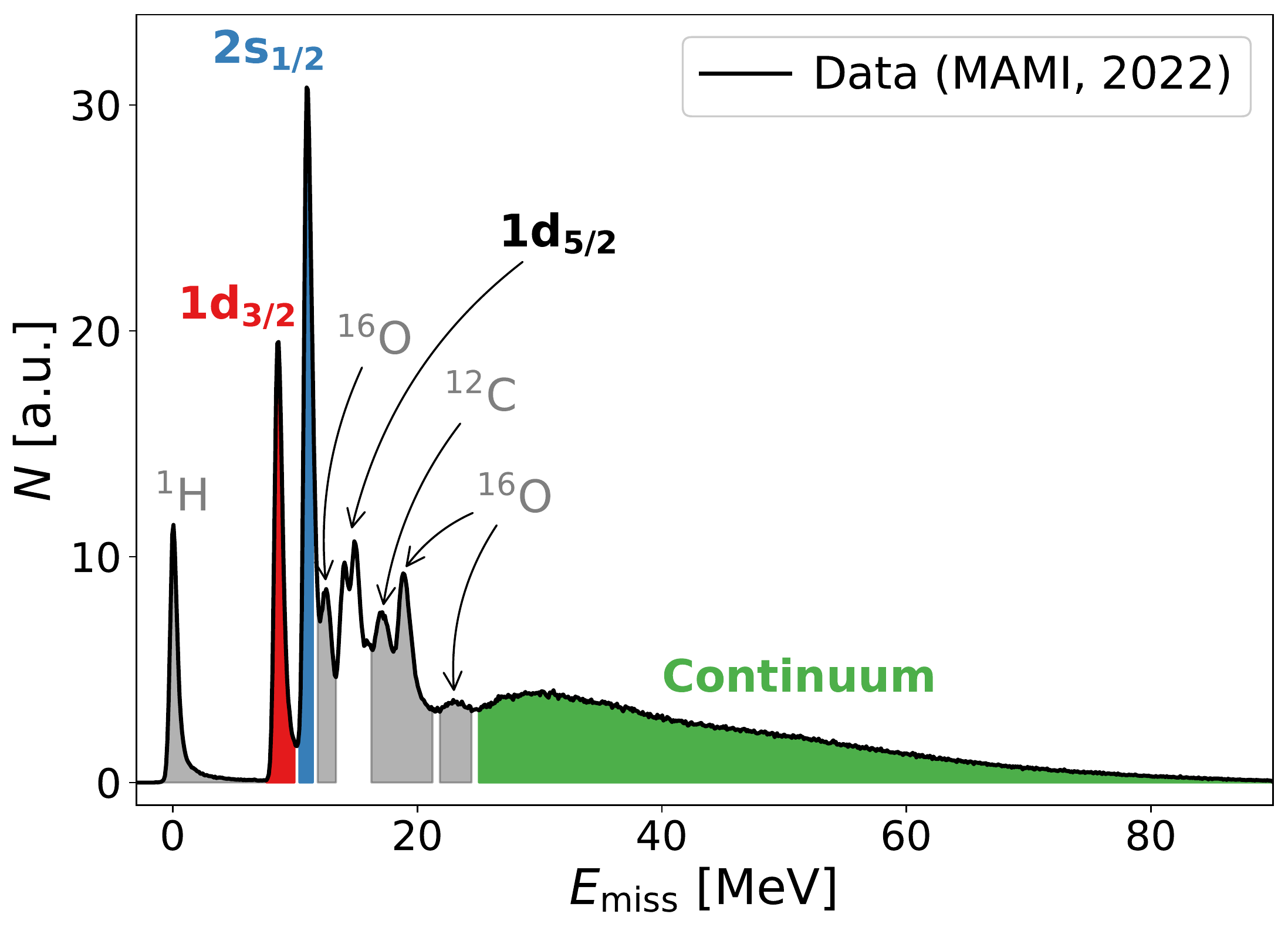}
\caption{
Missing-energy spectrum of the proton knockout from the $^{40}\mathrm{Ca}$ target. Marked are also peaks associated with contaminants from the residual target storage oil and oxidation.
}
\label{fig:emiss_histogram}
\end{figure}

The protons knocked out from the calcium shells were identified by their  missing energy defined as $E_{\rm miss}\equiv \omega - T_p - T_{{}^{39}{\rm K}}$. Here $\omega$ is the energy transfer, $T_p$ is the measured kinetic energy of the outgoing proton, and $T_{{}^{39}{\rm K}}$ is the calculated kinetic energy of the recoiling $^{39}{\rm K}$ nucleus. The $E_{\rm miss}$ spectrum shown in Fig.~\ref{fig:emiss_histogram} already includes the above-mentioned cuts. Well identified are removals where the residual system of $^{39}{\rm K}$ is left in the ground state ($J^{\pi}=\frac{3}{2}^{+}$) and in the first excited state at $E^{*}_{{}^{39}{\rm K}}=2.52\,\mathrm{MeV}$ ($J^{\pi}=\frac{1}{2}^{+}$). These two peaks correspond to protons removed from the $1\mathrm{d}_{3/2}$ and $2\mathrm{s}_{1/2}$ shells in $^{40}\mathrm{Ca}$, respectively, as was also confirmed by examining their $p_{\rm miss}$ distributions measured in the unpolarized $(e,e^{\prime}p)$ reaction~\cite{KRAMER1989199, KramerPhD, LAPIKAS1993297}. Events with missing energies above $25\,{\rm MeV}$ are attributed to the removal of the proton leaving the residual system in the continuum. 

We note that we do not resolve either $1\mathrm{f}_{7/2}$ ($2.814\,\mathrm{MeV}$) nor $2\mathrm{p}_{3/2}$ ($3.019\,\mathrm{MeV}$) that were observed in~\cite{KRAMER1989199, KramerPhD, LAPIKAS1993297}. However, with individual peaks having a FWHM around $0.6\,\mathrm{MeV}$, we can identify contributions to the measured spectrum from the various non-$^{40}\mathrm{Ca}$ contaminants; $^1{\rm H}$ and $^{12}{\rm C}$ from the target foil storage oil and $^{16}{\rm O}$ most likely due the oxidation during the transport of the foils to the target chamber. 

Having a clear elastic peak in the $E_{\rm miss}$ spectra enabled us to measure polarization transfer to a free proton. Being less than $1\sigma$ away from unity as shown later in Fig.~\ref{fig:virtDR}, the $^1\mathrm{H}$ result is consistent with the prediction for the polarization transfer in elastic scattering using $G_E/G_M$ from~\cite{Bernauer}. This indicates that the detector is performing well and provides further trust in our understanding of systematic uncertainties. 

To estimate systematic uncertainties we followed the same procedure as in~\cite{ceepTim}. Different contributions are summarized in Table~\ref{tab:systematics}. Beam polarization and analyzing power are the primary contributors to the uncertainty in the polarization components $P_x^{\prime}$ and $P_z^{\prime}$. However, their contribution largely cancel out when we form the ratio of the components. A limited resolution of the beam energy and spectrometers (central kinematics) both result in bin migration in target variables. Since VDCs and HDCs are relatively far apart, the quality of the alignment between the two is an important factor that contributes to the uncertainty in determining the secondary-scattering point and the related distributions. We also studied the contributions from various software cuts used in the analysis by slightly tightening each of them and observing the average effect of the modified cut over all the bins. We also evaluated the quality of the spin-precession calculation in our maximum-likelihood algorithm. The last item from Table~\ref{tab:systematics} relates to contamination of our continuum data with events coming from $^{12}{\rm C}$ and $^{16}{\rm O}$. We used both $(e,e')$ and $(e,e'p)$ missing energy distributions to estimate their contributions. The combined background from these two nuclei in the $^{40}\mathrm{Ca}$ continuum is around $10\%$. Based on moderate differences in polarization-transfer observables between different nuclei, we estimated $2\%$ as the upper limit on the uncertainty.

\begin{table}[ht]
\caption{
Contributions to the systematic uncertainties of the individual components, $P'_x$, $P'_z$, and their ratios, $(P'_x/P'_z)$. All values are in percent.}

\begin{tabular*}{\columnwidth}{l l @{\extracolsep{\fill}} r @{\extracolsep{16pt}} r @{\extracolsep{13pt}} r} 
\hline\hline\\[-6pt]
\multicolumn{2}{l}{} & $P'_x$  & $P'_z$ & $(P'_x/P'_z)$ \\[7pt]
\hline
\multicolumn{2}{l}{Beam pol.}               & 2.0    & 2.0 & $\approx$0.0    \hspace*{2mm}\\
\multicolumn{2}{l}{Analyzing power}         & 1.0    & 1.0 & $\approx$0.0    \hspace*{2mm}\\
\multicolumn{2}{l}{Beam energy}             & 0.2    & 0.6 & 0.8    \hspace*{2mm}\\
\multicolumn{2}{l}{Central kinematics}      & 0.6    & 0.9 & 1.0    \hspace*{2mm}\\
\multicolumn{2}{l}{HDC Alignment}           & $<$0.1 & 0.1 & 0.1    \hspace*{2mm}\\
\multicolumn{2}{l}{Software cuts}           & 1.6    & 1.7 & 1.5    \hspace*{2mm}\\
$E_{\rm miss}$ cut       & $1{\rm d}_{3/2}$ & 0.3    & 0.4 & 0.7    \hspace*{2mm}\\
                         & $2{\rm s}_{1/2}$ & 0.9    & 0.8 & 1.2    \hspace*{2mm}\\
                         & ${\rm Continuum}$& 0.4    & 0.7 & 0.9    \hspace*{2mm}\\
\multicolumn{2}{l}{Spin precession}    & 0.3    & 0.4 & $<$0.1    \hspace*{2mm}\\
\multicolumn{2}{l}{Contamination $(^{12}{\rm C},\,^{16}{\rm O})$}\\ 
                         & ${\rm Continuum}$ & 2.0    & 2.0 & 2.0    \hspace*{2mm}\\
\hline\hline
Total                    & $1{\rm d}_{3/2}$ & 2.9    & 3.1 & 2.1    \hspace*{2mm}\\
                         & $2{\rm s}_{1/2}$ & 3.0    & 3.1 & 2.3    \hspace*{2mm}\\
                         & ${\rm Continuum}$& 3.5    & 3.7 & 3.0    \hspace*{2mm}\\
\hline\hline
\end{tabular*}
\label{tab:systematics}
\end{table}

Due to the relative proximity of the two peaks related to $1\mathrm{d}_{3/2}$ and $2\mathrm{s}_{1/2}$, significant difference in their momentum distributions, and substantial radiative tail, we had to employ per-bin correction of polarization transfer to $2\mathrm{s}_{1/2}$ protons to account for the $1\mathrm{d}_{3/2}$ contribution. The correction was negligible for the two lowest $|p_{\rm miss}|$ bins. Thereafter, the share of protons coming from $1\mathrm{d}_{3/2}$ increased with the highest contribution of $15\%$. Because of very similar polarizations this correction was never larger than $2\%$. To further study the influence of $E_{\rm miss}$ cuts on the results, we varied each of the cuts independently by $\pm 0.25\,\mathrm{MeV}$ and added the deviations in quadrature. For the continuum the variation was increased to $\pm 5\,\mathrm{MeV}$ to account for possibility of several unresolved states in default cut's vicinity. We estimated that not resolving $1\mathrm{f}_{7/2}$ and $2\mathrm{p}_{3/2}$ below $2\mathrm{s}_{1/2}$ peak, results, within our $p_{\rm miss}$ range, in contamination on the order of few percent~\cite{KramerPhD}. Given our experience that the polarization transfer does not vary widely between states (and even nuclei), their contribution to polarization components is negligible.

\section{Polarization Transfer}
\subsection{Dependence on $\boldsymbol{p}_{\rm \mathbf{miss}}$}
\begin{figure}[ht!]
	\includegraphics[width=\columnwidth]{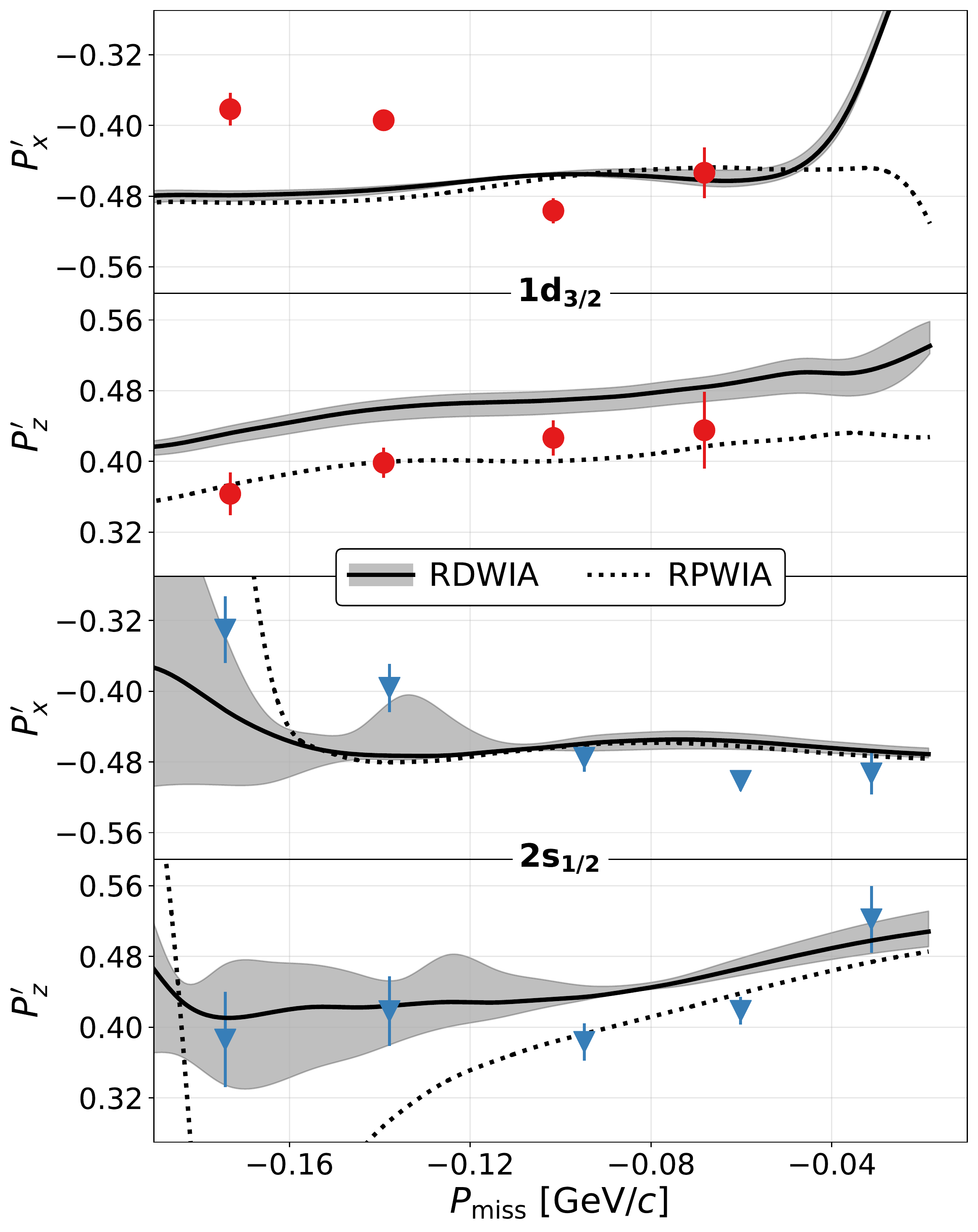}
	\caption{
 Polarization transfer components, $P_x^{\prime}$ and $P_z^{\prime}$, for protons knocked out of $1\mathrm{d}_{3/2}$ and $2\mathrm{s}_{1/2}$ shells of $^{40}\mathrm{Ca}$ nucleus. The errors are statistical only. Solid (dotted) lines ines are full RDWIA (RPWIA) calculations based on~\cite{Meucci:2001qc}. Bands around RDWIA calculations reflect variation due to use of different wavefunctions, optical potentials, and off-shell nuclear current prescriptions. See text for the description of the models. }
	\label{fig:40Ca_Components}
	\vspace{0cm}
\end{figure}

The components $P^{\prime}_x$ and $P^{\prime}_z$ of the polarization transfer to protons knocked out from $1\mathrm{d}_{3/2}$ and $2\mathrm{s}_{1/2}$ $^{40}\mathrm{Ca}$ shells are shown in Fig.~\ref{fig:40Ca_Components}. For the two discrete states we performed the RDWIA calculations (solid lines) using the global democratic optical potential from~\cite{Cooper:2009}, relativistic bound-state wave functions obtained with the NL-SH parametrization~\cite{SHARMA1993377}, and free-proton electromagnetic form factors from~\cite{Bernauer}. We adapted the original RDWIA program from~\cite{Meucci:2001qc} to include all 18 hadronic structure functions for the $\mathrm{A}(\vec{e}, e^{\prime} \vec{p})$ reaction in the Born approximation~\cite{Boffi:1996ikg}. The calculations were performed per-event using the individual measured kinematic parameters, fully matching the experimental kinematic acceptance. The deviations of the calculated components from the measured are significant for both shells. 

We show in Fig.~\ref{fig:40Ca_Components} the variations in RDWIA calculations (shaded regions) obtained with: i) two alternative parametrizations of the relativistic optical potentials; Energy-Dependent A-Independent fit to $^{40}\mathrm{Ca}$ data only and Energy-Dependent A-Dependent fit~\cite{Cooper:1993nx}, ii) another set of bound-state wavefunctions~\cite{PhysRevC.55.540}, and  iii) different off-shell nuclear current prescriptions~\cite{DEFOREST1983232}. Very little sensitivity to different ingredients is shown for knock out from the $1\mathrm{d}_{3/2}$. In case of $2\mathrm{s}_{1/2}$ shell, the same is true for the low $p_{\rm miss}$ region, while at the high $p_{\rm miss}$, the variation in both polarization components becomes relatively large. More details and individual calculations can be seen in Fig.~S1 of the supplementary material. RPWIA calculations analogous to the main RDWIA calculations are shwon as well. The main differences from the RDWIA calculations are found in the lognitudinal polarization component, $P_{z}^{\prime}$.

\begin{figure}[ht!]
	\includegraphics[width=\columnwidth]{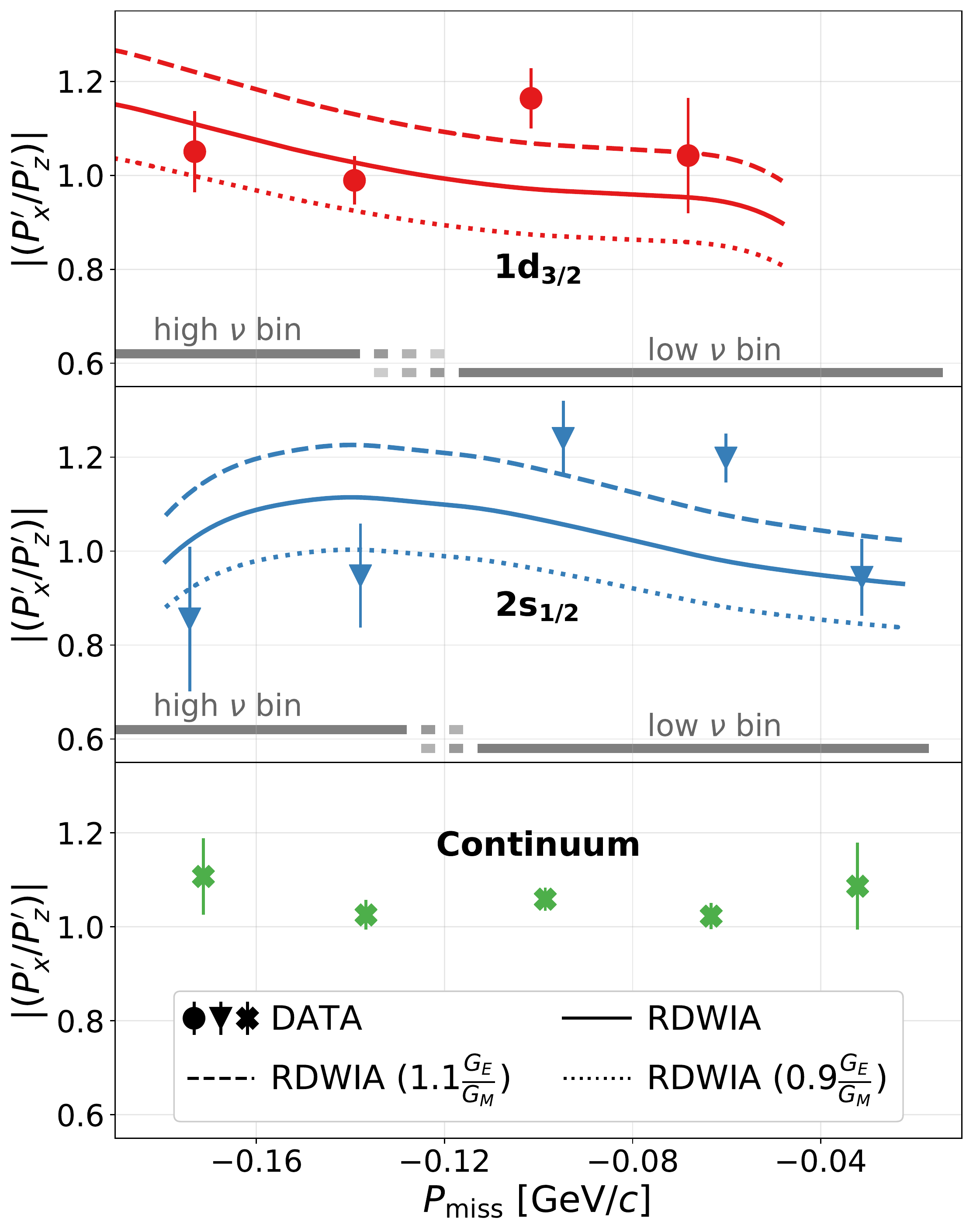}
	\caption{
 Ratios of polarization transfer components for protons from $1\mathrm{d}_{3/2}$ and $2\mathrm{s}_{1/2}$ shells as well as the continuum region. For the two discrete states the RDWIA calculation is shown, while there are no theory predictions for the continuum due to not well defined initial and final states. In addition dashed and dotted curves show RDWIA calculation with modifed $G_E/G_M$ by $+10\%$ and $-10\%$, respectively. At the bottom of the top two panels, the $p_{\rm miss}$ coverage of viruality ($\nu$) bins in Fig.~\ref{fig:virtDR} are shown. For the detailed relationship between $\nu$ and $p_{\rm miss}$ we refer the reader to Fig.~S2 of the supplementary material. }
	\label{fig:40Ca_SR}
	\vspace{0cm}
\end{figure}

\begin{figure*}[ht!]
	\includegraphics[width=\textwidth]{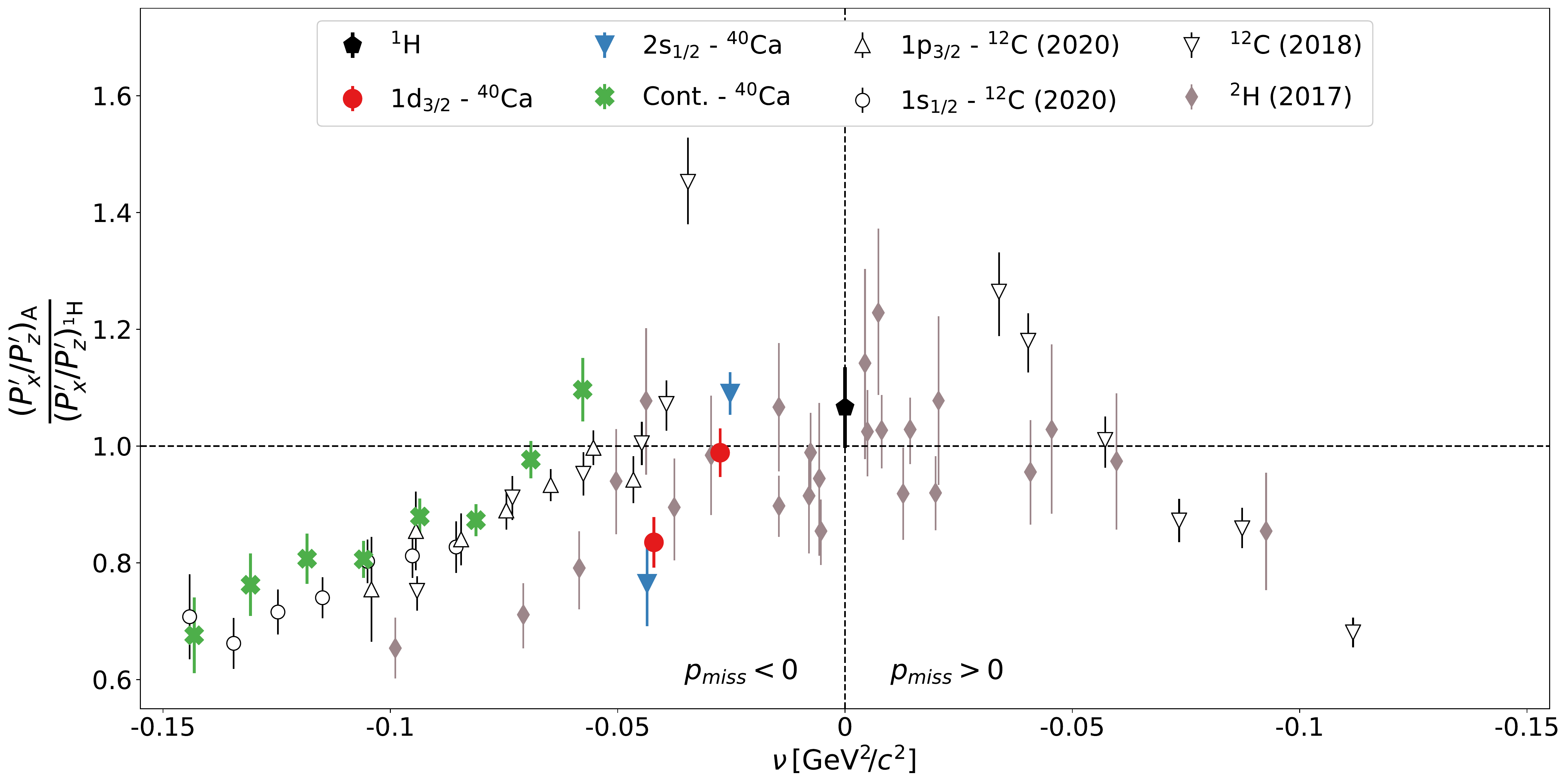}
	\caption{
		Ratio of measured polarization components divided by the calculated ratio for a moving free proton~\cite{Arenhovel} as a function of virtuality. Data of this work is compared to previously measured data on $^{2}\mathrm{H}$~\cite{deep2012PLB,deepCompPLB} and $^{12}\mathrm{C}$~\cite{ceepComp,ceepLet,ceepTim}.}
	\label{fig:virtDR}
	\vspace{0cm}
\end{figure*}

The ratios of the polarization-transfer components, $P^{\prime}_x/P^{\prime}_z$, are shown in Fig.~\ref{fig:40Ca_SR}, along with the RDWIA calculation (solid lines). We note that, compared to individual components, there is a better agreement between the calculated and measured ratios. The $P^{\prime}_x/P^{\prime}_z$ ratio depends linearly\footnote{In the elastic scattering under the Born approximation. In the quasi-elastic scattering in addition kinematics must be (anti)parallel and in-plane, as is the case in this measurement.} on the proton $G_E/G_M$ ratio. The sensitivity of the calculation to modification of the free-proton FF ratio was studied by repeating the calculations with free FF ratio modified by $\pm 10\%$. As can be seen, a change of roughly $10\%$ is comparable to the discrepancy between the measured and calculated (unmodified) ratio. This gives an estimate of the magnitude of shortcomings in the calculation, including possible changes in the proton EM form factors due to in-medium modifications of the bound nucleon.

The calculated components ratio was fitted to the data by adjusting $G_E/G_M$. For the $1\mathrm{d}_{3/2}$ protons the best fit value is $1.03\pm0.035$ ($\chi^2/\mathrm{DoF}=3.2$) which is consistent with no change. For the $2\mathrm{s}_{1/2}$ protons the best fit suggests a relatively large scaling factor of $1.11\pm0.035$ ($\chi^2/\mathrm{DoF}=3.9$). It seems that the differences between the calculations and data cannot be explained by a common $G_E/G_M$ scaling factor, and improvements are required at the level of addressing the individual components. Deviation of a similar size were observed for $^{4}\mathrm{He}$~\cite{Dieterich:2000mu,Strauch,Paolone} and were explained within a RDWIA framework by density-dependent $G_E$ and $G_M$ modifications from a quark-meson coupling model. An alternative explanation, within a non-relativistic framework, was proposed in terms of more conventional nuclear physics and includes many-body effects using accurate three- and four-nucleon bound-state wave functions, MEC, and isospin-dependent charge-exchange FSI~\cite{PhysRevLett.94.072303}.

For $E_{\rm miss}>25\,\mathrm{MeV}$, the residual system is in the continuum, and the polarization transfer represents an average of transfer to protons from various shells and no well-defined final state. This is the reason why the calculation is not available for those events. The data for the continuum has additional systematic uncertainty due to the target contamination with $^{12}\mathrm{C}$ and $^{16}\mathrm{O}$.

\subsection{Dependence on the bound-proton virtuality}
The polarization transfer ratio can also be characterized by the proton's virtuality, a measure of how far off-shell the proton is, defined as~\cite{deep2012PLB}:
\begin{equation}\label{eq:virtuality}
    \nu \equiv \left(M_{\rm A} - \sqrt{M^2_{\rm A-1}+p^2_{\rm miss}} \right)^2-p^2_{\rm miss}-M^2_{p}\,,
\end{equation}
where $M_{\rm A}$ is the mass of the target nucleus and $M_{\rm A-1}\equiv\sqrt{(\omega-E_{p}+M_{\rm A})^2 - p^2_{\rm miss}}$ is the residual mass. Notice that virtuality depends not only on $p_{\rm miss}$ but also $E_{\rm miss}$ indirectly through $M_{\rm A-1}$.

It has been observed that the deviation of the polarization transfer ratio of the knockout proton from that of a free proton $(P^{\prime}_x/P^{\prime}_z)_{\rm A}/(P^{\prime}_x/P^{\prime}_z)_{^{1}\mathrm{H}}$ in light nuclei, $^2{\rm H}$~\cite{deep2012PLB,deepCompPLB}, $^4{\rm He}$~\cite{Dieterich:2000mu,Strauch}, and $^{12}{\rm C}$~\cite{ceepComp,ceepTim} showed an overall agreement when compared at the same bound-proton virtuality as defined by Eq.~(\ref{eq:virtuality}). This agreement, together with the fact that these ratios were not only obtained for protons in different nuclei, but also measured at different kinematics, suggested that the deviation is best characterized by $\nu$. 

The double ratios from our measurement on $^{40}{\rm Ca}$ are shown in Fig.~\ref{fig:virtDR} along with previous determinations on $^{12}{\rm C}$ and $^2{\rm H}$. To compare quasi-elastic and elastic scattering, we accounted for the ``moving-proton'' correction that considers the Fermi motion of the struck proton~\cite{movingProtonPLB, ArenhovelMoving}. It was shown, however, that this affects the polarization ratio far less than the individual components.

The polarization transfer ratio to the protons associated with a residual system in the continuum follows the same pattern and shows good agreement with the light nuclei data. We note that these protons span over large $E_{\rm miss}$ range and can originate from various shells of $^{40}\mathrm{Ca}$. The measured ratios for protons ejected from the $1\mathrm{d}_{3/2}$ and $2\mathrm{s}_{1/2}$ shells cover a very small range in $\nu$, but seem to deviate from measurements done on $^{12}\mathrm{C}$ and the continuum $\mathrm{Ca}$ data. The well defined final state of the residual system in these events results in a very narrow $E_{\rm miss}$ range, which implies that the two bins in $\nu$ differ mainly in $p_{\rm miss}$ i.e. lower $\nu$ corresponds to lower $p_{\rm miss}$ (see gray bands in Fig.~\ref{fig:40Ca_SR} and Fig.~S2 of supplementary material). This is unlike in other data, where broader $E_{\rm miss}$ coverage can smear events with similar $p_{\rm miss}$ between several $\nu$ bins. This may indicate that the general agreement of the data results from averaging in each bin over large range in proton initial momentum. Phase space covered by data and more visual presentation of relation between $\nu$, $E_{\rm miss}$, and $p_{\rm miss}$ can be found in Fig.~S2 of the supplementary material. Future measurements extending to higher $p_{\rm miss}$ and/or in anti-parallel kinematics ($p_{\rm miss}>0$) will clarify the virtuality dependence of polarization transfer.

\section{Conclusions}
Our measurement of polarization transfer to a bound proton in $^{40}{\rm Ca}$ shows that while theoretical predictions are in general good agreement with the data, adjustments at the 5 to 10\% level are still needed. Interestingly, the polarization component ratio $P_x^{\prime}/P_z^{\prime}$ for both $1\mathrm{d}_{3/2}$ and $2\mathrm{s}_{1/2}$ states are underestimated, which may point to a common deficiency in the calculations. 

The virtuality dependence of the double ratio of the $^{40}\mathrm{Ca}$ components to those of a free proton seem to agree with data obtained for $^{12}\mathrm{C}$ and $^{2}\mathrm{H}$, for the continuum states, but may show a deviation for the transitions to a discrete state in the residual nucleus. It may point to the fact that for such states the $E_{\rm miss}$ range in the virtuality bin is very narrow, resulting in almost no $p_{\rm miss}$ overlap between different $\nu$ bins. This is different for the continuum which spans a relatively large $E_{\rm miss}$ and generally wider $p_{\rm miss}$ range. This dependence should be further studied by extending the kinematic range of the measurement.

\section{Acknowledgements}
We would like to thank the Mainz Microtron operators and technical crew for the excellent operation of the accelerator during the challenging times of worldwide pandemic. This work is supported by the Israel Science Foundation (Grant 951/19) of the Israel Academy of Arts and Sciences, by the Israel Ministry of Science, Technology and Space, by the PAZY Foundation (Grant 294/18), by the the PRISMA+ (Precision Physics, Fundamental Interactions and Structure of Matter) Cluster of Excellence, by the Deutsche Forschungsgemeinschaft (Collaborative Research Center 1044), by the Federal State of Rhineland-Palatinate, by the U.S. National Science Foundation (PHY-2111050), and by the  United States-Israeli Binational Science Foundation (BSF) as part of the joint program with the NSF (grant 2020742). We also acknowledge the financial support from the Slovenian Research Agency (research core funding No.~P1\textendash 0102) and the Research Excellence Initiative of the University of Silesia in Katowice.  
\FloatBarrier
\bibliographystyle{elsarticle-num}

\addcontentsline{toc}{section}{\refname}\small{\bibliography{hdep}}

\begin{thebibliography}{10}
\expandafter\ifx\csname url\endcsname\relax
  \def\url#1{\texttt{#1}}\fi
\expandafter\ifx\csname urlprefix\endcsname\relax\def\urlprefix{URL }\fi
\expandafter\ifx\csname href\endcsname\relax
  \def\href#1#2{#2} \def\path#1{#1}\fi

\bibitem{PhysRevC.31.232}
L.~S. Celenza, A.~Rosenthal, C.~M. Shakin, Many-body soliton dynamics:
  Modification of nucleon properties in nuclei, Phys. Rev. C 31 (1985)
  232--239.
\newblock \href {https://doi.org/10.1103/PhysRevC.31.232}
  {\path{doi:10.1103/PhysRevC.31.232}}.

\bibitem{BERGMANN1990185}
M.~Bergmann, K.~Goeke, S.~Krewald, Medium effects in quasi-elastic electron
  scattering, Physics Letters B 243~(3) (1990) 185--190.
\newblock \href {https://doi.org/https://doi.org/10.1016/0370-2693(90)90837-V}
  {\path{doi:https://doi.org/10.1016/0370-2693(90)90837-V}}.

\bibitem{LU1998217}
D.~Lu, A.~Thomas, K.~Tsushima, A.~Williams, K.~Saito, In-medium
  electron-nucleon scattering, Physics Letters B 417~(3) (1998) 217--223.
\newblock \href {https://doi.org/https://doi.org/10.1016/S0370-2693(97)01385-3}
  {\path{doi:https://doi.org/10.1016/S0370-2693(97)01385-3}}.

\bibitem{PhysRevC.60.068201}
D.~H. Lu, K.~Tsushima, A.~W. Thomas, A.~G. Williams, K.~Saito, Electromagnetic
  form factors of the bound nucleon, Phys. Rev. C 60 (1999) 068201.
\newblock \href {https://doi.org/10.1103/PhysRevC.60.068201}
  {\path{doi:10.1103/PhysRevC.60.068201}}.

\bibitem{PhysRevC.76.055206}
C.~C.~d. Atti, L.~L. Frankfurt, L.~P. Kaptari, M.~I. Strikman, Dependence of
  the wave function of a bound nucleon on its momentum and the emc effect,
  Phys. Rev. C 76 (2007) 055206.
\newblock \href {https://doi.org/10.1103/PhysRevC.76.055206}
  {\path{doi:10.1103/PhysRevC.76.055206}}.

\bibitem{PhysRevC.87.028202}
G.~Ron, W.~Cosyn, E.~Piasetzky, J.~Ryckebusch, J.~Lichtenstadt, Nuclear density
  dependence of in-medium polarization, Phys. Rev. C 87 (2013) 028202.
\newblock \href {https://doi.org/10.1103/PhysRevC.87.028202}
  {\path{doi:10.1103/PhysRevC.87.028202}}.

\bibitem{Akh74}
A.~I. Akhiezer, M.~Rekalo,
  \href{http://refhub.elsevier.com/S0370-2693(17)30052-7/bib416B683734s1}{Polarization
  effects in the scattering of leptons by hadrons}, Sov. J. Part. Nucl. 4
  (1974) 277, [Fiz. Elem. Chast. Atom. Yadra 4, (1973) 662].

\bibitem{PhysRevC.40.290}
A.~Picklesimer, J.~W. Van~Orden, Polarization response functions and the
  $(\vec{e},e^{\prime}\vec{p}\,)$ reaction, Phys. Rev. C 40 (1989) 290--303.
\newblock \href {https://doi.org/10.1103/PhysRevC.40.290}
  {\path{doi:10.1103/PhysRevC.40.290}}.

\bibitem{Kelly:1996hd}
J.~J. Kelly, {Nucleon knockout by intermediate-energy electrons}, Adv. Nucl.
  Phys. 23 (1996) 75--294.
\newblock \href {https://doi.org/10.1007/0-306-47067-5_2}
  {\path{doi:10.1007/0-306-47067-5_2}}.

\bibitem{Milbrath:1997de}
B.~D. Milbrath, J.~I. McIntyre, et~al., {A Comparison of polarization
  observables in electron scattering from the proton and deuteron}, Phys. Rev.
  Lett. 80 (1998) 452--455, [Erratum: Phys. Rev. Lett. 82, 2221 (1999)].
\newblock \href {http://arxiv.org/abs/nucl-ex/9712006}
  {\path{arXiv:nucl-ex/9712006}}, \href
  {https://doi.org/10.1103/PhysRevLett.80.452, 10.1103/PhysRevLett.82.2221}
  {\path{doi:10.1103/PhysRevLett.80.452, 10.1103/PhysRevLett.82.2221}}.

\bibitem{PhysRevC.73.064004}
B.~Hu, M.~K. Jones, P.~E. Ulmer, et~al., Polarization transfer in the
  $^{2}\mathrm{H}(\vec{e},{e}^{\prime}\vec{p})n$ reaction up to
  ${Q}^{2}=1.61\phantom{\rule{0.3em}{0ex}}(\mathrm{GeV}/c){}^{2}$, Phys. Rev. C
  73 (2006) 064004.
\newblock \href {https://doi.org/10.1103/PhysRevC.73.064004}
  {\path{doi:10.1103/PhysRevC.73.064004}}.

\bibitem{deep2012PLB}
I.~Yaron, D.~Izraeli, et~al., Polarization-transfer measurement to a
  large-virtuality bound proton in the deuteron, Phys.\ Lett.\ B 769 (2017)
  21--24.
\newblock \href {https://doi.org/10.1016/j.physletb.2017.01.034}
  {\path{doi:10.1016/j.physletb.2017.01.034}}.

\bibitem{deepCompPLB}
D.~Izraeli, I.~Yaron, et~al., {Components of polarization-transfer to a bound
  proton in a deuteron measured by quasi-elastic electron scattering}, Phys.
  Lett. B 781 (2018) 107--111.
\newblock \href {http://arxiv.org/abs/1801.01306} {\path{arXiv:1801.01306}},
  \href {https://doi.org/10.1016/j.physletb.2018.03.063}
  {\path{doi:10.1016/j.physletb.2018.03.063}}.

\bibitem{deepPaul}
S.~Paul, D.~Izraeli, T.~Brecelj, I.~Yaron, et~al., {Polarization-transfer
  measurements in deuteron quasi-elastic anti-parallel kinematics}, Phys. Lett.
  B 795C (2019) 599--605.
\newblock \href {http://arxiv.org/abs/1905.05594} {\path{arXiv:1905.05594}},
  \href {https://doi.org/10.1016/j.physletb.2019.07.002}
  {\path{doi:10.1016/j.physletb.2019.07.002}}.

\bibitem{Dieterich:2000mu}
S.~Dieterich, et~al., {Polarization transfer in the $^4$He$(\vec e, e' \vec
  p)^3$H reaction}, Phys. Lett. B 500 (2001) 47--52.
\newblock \href {http://arxiv.org/abs/nucl-ex/0011008}
  {\path{arXiv:nucl-ex/0011008}}, \href
  {https://doi.org/10.1016/S0370-2693(01)00052-1}
  {\path{doi:10.1016/S0370-2693(01)00052-1}}.

\bibitem{Strauch}
S.~Strauch, et~al., {Polarization transfer in the ${^4}$\textsc{H}e$(\vec
  e,e'\vec p)$${^3}$\textsc{H} reaction up to Q${^2}$ = 2.6~(GeV/$c$)${^2}$},
  Phys. Rev. Lett. 91 (2003) 052301.
\newblock \href {https://doi.org/10.1103/PhysRevLett.91.052301}
  {\path{doi:10.1103/PhysRevLett.91.052301}}.

\bibitem{Paolone}
M.~Paolone, S.~P. Malace, S.~Strauch, et~al., Polarization transfer in the
  $^{4}\mathrm{He}(\vec e,e'\vec{p}\,)^{3}\mathrm{H}$ reaction at ${Q}^{2}=0.8$
  and $1.3\,(\mathrm{GeV}/c{)}^{2}$, Phys. Rev. Lett. 105 (2010) 072001.
\newblock \href {https://doi.org/10.1103/PhysRevLett.105.072001}
  {\path{doi:10.1103/PhysRevLett.105.072001}}.

\bibitem{ceepLet}
D.~Izraeli, T.~Brecelj, et~al., {Measurement of polarization-transfer to bound
  protons in carbon and its virtuality dependence}, Phys. Lett. B 781 (2018)
  95--98.
\newblock \href {http://arxiv.org/abs/1711.09680} {\path{arXiv:1711.09680}},
  \href {https://doi.org/10.1016/j.physletb.2018.03.027}
  {\path{doi:10.1016/j.physletb.2018.03.027}}.

\bibitem{ceepComp}
T.~Brecelj, S.~J. Paul, T.~Kolar, et~al., Polarization transfer to bound
  protons measured by quasielastic electron scattering on $^{12}\mathrm{C}$,
  Phys. Rev. C 101 (2020) 064615.
\newblock \href {https://doi.org/10.1103/PhysRevC.101.064615}
  {\path{doi:10.1103/PhysRevC.101.064615}}.

\bibitem{ceepTim}
T.~Kolar, S.~Paul, T.~Brecelj, et~al., {Comparison of recoil polarization in
  the $^{12}{\rm C}(\vec e,e'\vec p)$ process for protons extracted from $s$
  and $p$ shell}, Phys. Lett. B 811 (2020) 135903.
\newblock \href {http://arxiv.org/abs/2007.14985} {\path{arXiv:2007.14985}},
  \href {https://doi.org/10.1016/j.physletb.2020.135903}
  {\path{doi:10.1016/j.physletb.2020.135903}}.

\bibitem{Malov_O16}
S.~Malov, et~al., Polarization transfer in the $^{16}\mathrm{O}(\vec e,e'\vec
  p)^{15}\mathrm{N}$ reaction, Phys.\ Rev.\ C 62 (2000) 057302.
\newblock \href {https://doi.org/10.1103/PhysRevC.62.057302}
  {\path{doi:10.1103/PhysRevC.62.057302}}.

\bibitem{LAPIKAS1993297}
L.~Lapik\'{a}s, Quasi-elastic electron scattering off nuclei, Nuclear Physics A
  553 (1993) 297--308.
\newblock \href {https://doi.org/https://doi.org/10.1016/0375-9474(93)90630-G}
  {\path{doi:https://doi.org/10.1016/0375-9474(93)90630-G}}.

\bibitem{KRAMER1989199}
G.~Kramer, et~al., Proton ground-state correlations in $^{40}{Ca}$ studied with
  the reaction $^{40}\mathrm{Ca}(e,e^{\prime}p){}^{39}\mathrm{K}$, Physics
  Letters B 227~(2) (1989) 199--203.
\newblock \href {https://doi.org/https://doi.org/10.1016/S0370-2693(89)80022-X}
  {\path{doi:https://doi.org/10.1016/S0370-2693(89)80022-X}}.

\bibitem{KramerPhD}
G.~Kramer,
  \href{https://inis.iaea.org/collection/NCLCollectionStore/_Public/22/024/22024992.pdf}{{The
  proton spectral function of $^{40}{Ca}$ and $^{48}{Ca}$ studied with the
  $(e,e^{\prime}p)$ reaction: An Investigation of Ground-state Correlations.}},
  Ph.D. thesis, {Vrije Universiteit, Amsterdam.} (1990).
\newline\urlprefix\url{https://inis.iaea.org/collection/NCLCollectionStore/_Public/22/024/22024992.pdf}

\bibitem{a1aparatus}
K.~Blomqvist, et~al., The three-spectrometer facility at {MAMI}, Nucl.\
  Instrum.\ and Meth.\ A 403~(2--3) (1998) 263 -- 301.
\newblock \href {https://doi.org/10.1016/S0168-9002(97)01133-9}
  {\path{doi:10.1016/S0168-9002(97)01133-9}}.

\bibitem{Pospischil:2000pu}
T.~Pospischil, et~al., The focal plane proton-polarimeter for the
  3-spectrometer setup at {MAMI}, Nucl.\ Instrum.\ Methods.\ Phys.\ Res.,
  Sect.\ A 483~(3) (2002) 713 -- 725.
\newblock \href {https://doi.org/10.1016/S0168-9002(01)01955-6}
  {\path{doi:10.1016/S0168-9002(01)01955-6}}.

\bibitem{Bernauer}
J.~C. Bernauer, M.~O. Distler, J.~Friedrich, T.~Walcher, P.~Achenbach,
  C.~Ayerbe-Gayoso, et~al., {Electric and magnetic form factors of the proton},
  Phys.\ Rev. C 90~(1) (2014) 015206.
\newblock \href {https://doi.org/10.1103/PhysRevC.90.015206}
  {\path{doi:10.1103/PhysRevC.90.015206}}.

\bibitem{Cooper:2009}
E.~D. Cooper, S.~Hama, B.~C. Clark, Global dirac optical potential from helium
  to lead, Phys. Rev. C 80 (2009) 034605.
\newblock \href {https://doi.org/10.1103/PhysRevC.80.034605}
  {\path{doi:10.1103/PhysRevC.80.034605}}.

\bibitem{SHARMA1993377}
M.~Sharma, M.~Nagarajan, P.~Ring, Rho meson coupling in the relativistic mean
  field theory and description of exotic nuclei, Phys. Lett. B 312~(4) (1993)
  377 -- 381.
\newblock \href {https://doi.org/https://doi.org/10.1016/0370-2693(93)90970-S}
  {\path{doi:https://doi.org/10.1016/0370-2693(93)90970-S}}.

\bibitem{Meucci:2001qc}
A.~Meucci, C.~Giusti, F.~D. Pacati, {Relativistic corrections in $(e,
  e^{\prime} p)$ knockout reactions}, Phys. Rev. C 64 (2001) 014604.
\newblock \href {http://arxiv.org/abs/nucl-th/0101034}
  {\path{arXiv:nucl-th/0101034}}, \href
  {https://doi.org/10.1103/PhysRevC.64.014604}
  {\path{doi:10.1103/PhysRevC.64.014604}}.

\bibitem{Boffi:1996ikg}
S.~Boffi, C.~Giusti, F.~d. Pacati, M.~Radici, {Electromagnetic Response of
  Atomic Nuclei}, Vol.~20 of Oxford Studies in Nuclear Physics, Clarendon
  Press, Oxford UK, 1996.

\bibitem{Cooper:1993nx}
E.~D. Cooper, S.~Hama, B.~C. Clark, R.~L. Mercer, {Global Dirac phenomenology
  for proton nucleus elastic scattering}, Phys. Rev. C 47 (1993) 297--311.
\newblock \href {https://doi.org/10.1103/PhysRevC.47.297}
  {\path{doi:10.1103/PhysRevC.47.297}}.

\bibitem{PhysRevC.55.540}
G.~A. Lalazissis, J.~K\"onig, P.~Ring, New parametrization for the lagrangian
  density of relativistic mean field theory, Phys. Rev. C 55 (1997) 540--543.
\newblock \href {https://doi.org/10.1103/PhysRevC.55.540}
  {\path{doi:10.1103/PhysRevC.55.540}}.

\bibitem{DEFOREST1983232}
T.~{De Forest}, Off-shell electron-nucleon cross sections: The impulse
  approximation, Nucl. Phys. A 392~(2) (1983) 232 -- 248.
\newblock \href {https://doi.org/https://doi.org/10.1016/0375-9474(83)90124-0}
  {\path{doi:https://doi.org/10.1016/0375-9474(83)90124-0}}.

\bibitem{Arenhovel}
H.~Arenh{\"o}vel, W.~Leidemann, E.~L. Tomusiak, {General survey of polarization
  observables in deuteron electrodisintegration}, Eur.\ Phys.\ J. A 23 (2005)
  147--190.
\newblock \href {https://doi.org/10.1140/epja/i2004-10061-5}
  {\path{doi:10.1140/epja/i2004-10061-5}}.

\bibitem{PhysRevLett.94.072303}
R.~Schiavilla, O.~Benhar, A.~Kievsky, L.~E. Marcucci, M.~Viviani, Polarization
  transfer in $^{4}\mathrm{He}(\vec e,e'\vec p\,)^{3}\mathrm{H}$: Is the ratio
  ${G}_{Ep}/{G}_{Mp}$ modified in the nuclear medium?, Phys. Rev. Lett. 94
  (2005) 072303.
\newblock \href {https://doi.org/10.1103/PhysRevLett.94.072303}
  {\path{doi:10.1103/PhysRevLett.94.072303}}.

\bibitem{movingProtonPLB}
S.~Paul, T.~Brecelj, H.~Arenh{\"o}vel, et~al., {The influence of Fermi motion
  on the comparison of the polarization transfer to a proton in elastic $\vec
  ep$ and quasi-elastic $\vec eA$ scattering}, Phys. Lett. B 792 (2019)
  445--449.
\newblock \href {http://arxiv.org/abs/1901.10958} {\path{arXiv:1901.10958}},
  \href {https://doi.org/10.1016/j.physletb.2019.04.004}
  {\path{doi:10.1016/j.physletb.2019.04.004}}.

\bibitem{ArenhovelMoving}
H.~Arenh{\"o}vel, {Polarization observables for elastic electron scattering off
  a moving nucleon}, Phys. Rev. C 99 (2019) 055502.
\newblock \href {http://arxiv.org/abs/1904.04515} {\path{arXiv:1904.04515}},
  \href {https://doi.org/10.1103/PhysRevC.99.055502}
  {\path{doi:10.1103/PhysRevC.99.055502}}.

\end{thebibliography}
\clearpage

\end{document}